\documentclass[aps,pra,reprint,superscriptaddress]{revtex4-1}

\usepackage{graphicx}
\usepackage{multirow}

\begin{document}

\title[]{High-resolution mid-infrared spectroscopy of buffer-gas-cooled methyltrioxorhenium molecules}%

\author{S. K. Tokunaga}
\affiliation{These authors contributed equally to this work.}
\affiliation{Universit\'{e} Paris 13, Sorbonne Paris Cit\'{e}, Laboratoire de Physique des Lasers, F-93430 Villetaneuse, France}
\affiliation{CNRS, UMR 7538, LPL, F-93430 Villetaneuse, France}

\author{R. J. Hendricks}
\altaffiliation[Present address: ]{National Physical Laboratory, Hampton Road, Teddington, TW11 0LW, UK}
\affiliation{These authors contributed equally to this work.}
\affiliation{Centre for Cold Matter, Blackett Laboratory, Imperial College London, Prince Consort Road, London SW7 2AZ, UK}

\author{M. R. Tarbutt}
\email{m.tarbutt@imperial.ac.uk}
\affiliation{Centre for Cold Matter, Blackett Laboratory, Imperial College London, Prince Consort Road, London SW7 2AZ, UK}
\author{B. Darqui\'e}
\email{benoit.darquie@univ-paris13.fr}
\affiliation{CNRS, UMR 7538, LPL, F-93430 Villetaneuse, France}
\affiliation{Universit\'{e} Paris 13, Sorbonne Paris Cit\'{e}, Laboratoire de Physique des Lasers, F-93430 Villetaneuse, France}

\begin{abstract}
We demonstrate cryogenic buffer-gas cooling of gas-phase methyltrioxorhenium (MTO). This molecule is closely related to chiral organometallic molecules where the parity-violating energy differences between enantiomers may be measurable. The molecules are produced with a rotational temperature of approximately 6~K by laser ablation of an MTO pellet inside a cryogenic helium buffer gas cell. Facilitated by the low temperature, we demonstrate absorption spectroscopy of the 10.2~$\mu$m antisymmetric Re=O stretching mode of MTO with a resolution of 8~MHz and a frequency accuracy of 30~MHz. We partially resolve the hyperfine structure and measure the nuclear quadrupole coupling of the excited vibrational state.
\end{abstract}

\maketitle

\section{Introduction}

There is a demand for precise spectroscopy of increasingly complex molecules. Precise knowledge of molecular constants is required for modelling our atmosphere\cite{Guinet2010,Harrison2011}, for the interpretation of spectra of astrophysical and planetary interest\cite{Coudert2006}, for studying collision physics\cite{Hartmann2008,Buldyreva2011}, and for performing tests of fundamental physics, ranging from searches for signatures of symmetry breaking\cite{Daussy1999,Hudson2011,Tokunaga2013,Baron2014,Cahn2014} to measurements of the fundamental constants\cite{Mejri2015,Biesheuvel2016} and their possible variation\cite{Hudson2006,Shelkovnikov2008,Truppe2013,Jansen2014}.

One such test is the measurement of energy differences between chiral enantiomers induced by parity violation. The idea that parity violation might be manifest in chiral species dates back to the 1970s\cite{Rein1974,Letokhov1975}. A precise measurement could contribute to the determination of several parameters of the standard model of particle physics\cite{Bouchiat1997}, provide calibrations of relativistic quantum chemistry calculations\cite{Darquie2010,Quack2008}, and shed light on the question of biomolecular homochirality\cite{Viedma2011}. We are constructing a Ramsey interferometer to compare the vibrational frequencies of the two enantiomers of a chiral molecule\cite{Darquie2010, Tokunaga2013}. For this work, a cold, slow-moving, intense beam of these molecules is desirable to increase the resolution, the statistical precision and to reduce systematic uncertainties relating to the beam velocity.  Buffer-gas-cooled beams, developed and comprehensively reviewed\cite{Hutzler2012} by the group of J. Doyle at Harvard, can provide the low temperature, low speed, and high intensity needed\cite{Hutzler2011,Hutzler2012}.

To form a buffer-gas-cooled beam, hot molecules are introduced into a cell containing helium at $\sim$4~K, where they cool through collisions with the helium and then flow out of an exit hole. Various methods have been demonstrated for loading molecules into the cell, including capillary injection\cite{Patterson2007}, direct loading from an oven or beam\cite{Patterson2015}, and laser ablation\cite{Hutzler2011, Barry2011, Bulleid2013}. The methods were first developed for diatomic molecules, but more recently buffer-gas cooling of complex molecules has been demonstrated. Organic molecules with high volatility have been introduced via direct gas flow\cite{Patterson2010} or from an oven or beam\cite{Patterson2012,Patterson2013,Patterson2013b,Piskorski2014,Spaun2016}. Various spectroscopic techniques have been applied to these cooled molecules, taking advantage of the narrow linewidths, larger signals and simplified spectra that result from the cooling. These techniques include Fourier transform microwave spectroscopy\cite{Patterson2012,Patterson2013,Patterson2013b,Patterson2015}, visible and ultra-violet light induced fluorescence\cite{Patterson2010,Piskorski2014}, and cavity-enhanced frequency comb spectroscopy in the $3-5~\mu$m spectral region\cite{Spaun2016}. These techniques have enabled the exploration of molecule-molecule collisions and the inhibition of clustering between the molecules and the buffer gas atoms\cite{Patterson2010,Patterson2012,Piskorski2014,Spaun2016}, measurements of previously inaccessible low temperature rotational and vibrational relaxation cross-sections\cite{Patterson2012}, the elucidation of rovibrational structures and the determination of molecular parameters\cite{Spaun2016}, and the demonstration of a novel chiral analysis method for measuring enantiomeric excess\cite{Patterson2013,Patterson2013b}.

In this paper, we report the buffer-gas cooling of methyltrioxorhenium (MTO, CH$_3$ReO$_3$, a far less volatile complex molecule and an achiral precursor of molecules of interest for parity violation measurements in chiral molecules\cite{Saleh2013}. We introduce the molecules into the cell by laser ablation, showing that the organometallic species of interest can survive the ablation process, and that they are well-suited to the generation of buffer-gas beams. We also demonstrate the first precise spectroscopic measurements of buffer-gas-cooled molecules in the mid-infrared region around $10~\mu$m, obtaining rotational and hyperfine-resolved absorption spectra in the Re=O stretching region of MTO. Our work extends the cooling technique to a new class of molecules and demonstrates the potential for high resolution spectroscopic measurements of these cooled molecules in the fingerprint region. We note that, in a recent experiment at Harvard, MTO has been injected into a cryogenic buffer-gas cell from a heated pulsed valve, and broadband Fourier transform microwave spectroscopy of the molecules was performed\cite{Drayna2016}.

\section{Apparatus}

\begin{figure}
 \centering
 \includegraphics[width=\columnwidth]{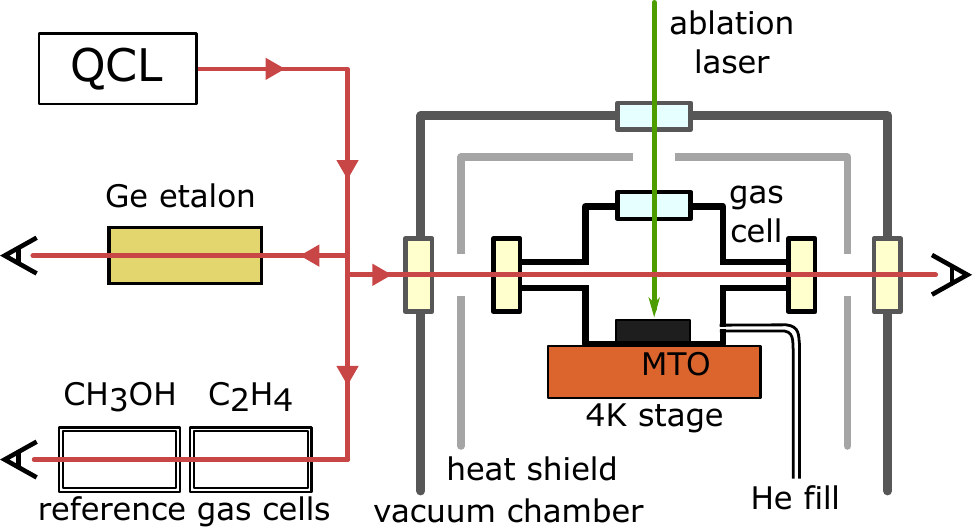}
 \caption{The experimental setup. QCL: quantum cascade laser. MTO: methyltrioxorhenium pellet.}
 \label{fgr:setup}
\end{figure}

The experimental setup is shown schematically in figure~\ref{fgr:setup}. The cryogenic apparatus, described in more detail elsewhere\cite{Bulleid2013}, consists of a closed copper cell with a cubic internal void of 30~mm side length attached to the cold stage of a closed-cycle cryocooler (Sumitomo RDK-415D).  The cell has a measured temperature of 6~K.  Windows are indium-sealed onto three sides of the cell to provide optical access, and a narrow tube is used to fill the cell with helium gas.  We find that the ablation process (see below) alters the helium pressure in the cell. During operation, the helium pressure measured at the room-temperature end of the tube is 1$\times 10^{-2}$~mbar, which corresponds to a time-averaged helium density inside the cell of approximately $4\times 10^{15}$~cm$^{-3}$.  The cell is surrounded by aluminium radiation shields at a temperature of about 40~K, and is housed in a vacuum chamber at a pressure of $4\times 10^{-7}$~mbar.

Inside the cell, mounted on one face, is a pellet made from 98\% purity MTO (Strem Chemical Inc.) that has been ground in a mortar, mixed with a few percent of graphite to help make it opaque to light at 1064~nm, and then pressed for several minutes in a 2-ton pneumatic press.  MTO molecules are injected into the cold helium buffer gas by ablating this pellet with 30~mJ pulses from a Nd:YAG laser (Quantel Ultra 50) operating at 1064~nm.  The pulses have a duration of 8~ns and the beam is focussed to a spot size which we estimate to be 150~$\mathrm{\mu}$m.

MTO molecules liberated from the pellet thermalise with the helium buffer gas and are detected by laser absorption spectroscopy of the Re=O antisymmetric stretching mode at about 976~cm${}^{-1}$.  The probe laser is a free-running continuous-wave single-mode distributed-feedback quantum cascade laser (QCL; Alpes Lasers) with an estimated linewidth of about 1~MHz at 1~s (which is the typical timescale for a measurement at a single frequency; see later). When operated at a temperature of 263~K it delivers about 40~mW of light at $\sim10.25~\mu$m. About 1~$\mu$W of this light, in a beam with a diameter of $\approx 3$~mm, is sent through the cell via Zn:Se windows mounted on stand off tubes to prevent contamination by ablated material. The transmitted power is measured on a liquid nitrogen cooled HgCdTe detector (Teledyne Judson Technologies, J19D11-M209-R250U-30).  The laser frequency can be changed by stepping the supply current.  The relative frequency is monitored by measuring the intensity transmitted through one of a pair of temperature-stabilised solid germanium etalons with free spectral ranges of 1450 and 490~MHz.  Absolute laser frequencies can be obtained at the same time by measuring the transmission through a pair of reference cells containing methanol and ethene respectively at a pressure of approximately 3~mbar.  Both of these species have been extensively studied and have accurate absorption line frequencies listed in the HITRAN database\cite{Rothman2013,HITRANwebsite}.

\section{Results}

\begin{figure}
 \centering
 \includegraphics[width=1\columnwidth]{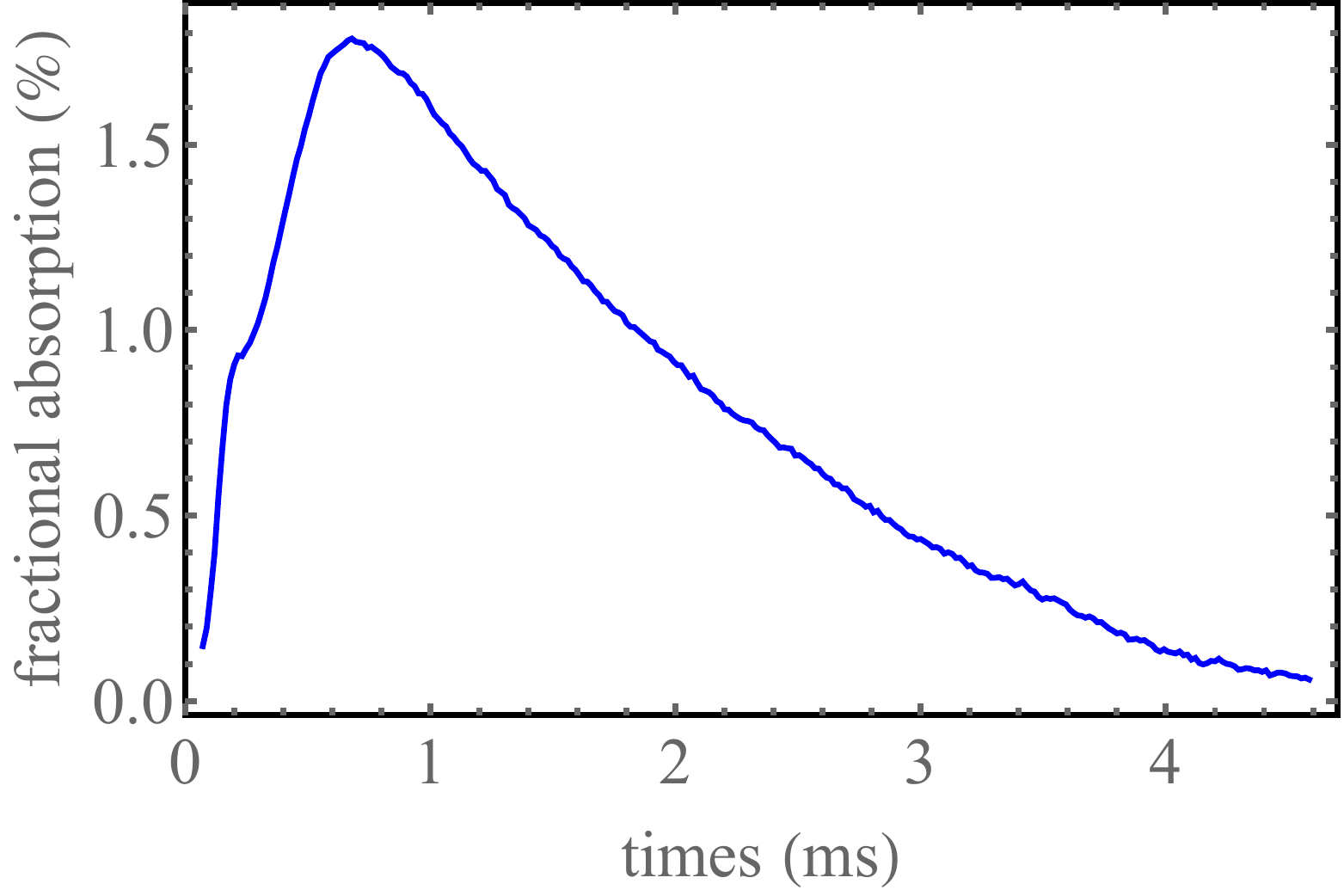}
 \caption{Laser absorption in the cell as a function of time after the ablation pulse, which defines $t=0$. Average of 25 consecutive shots.}
 \label{fgr:TOF}
\end{figure}

In each `shot' of the experiment the probe laser has a fixed frequency and the intensities transmitted through the cryogenic cell, the etalon and the reference gas cells are recorded as a function of time $t$ for 10~ms after an ablation pulse. Figure~\ref{fgr:TOF} shows MTO absorption as a function of time, averaged over 25 successive shots.  With the laser on resonance with a strong transition, and under optimised conditions, we see a rapid rise in absorption to a peak of a few percent followed by a slower, roughly exponential, decay.  We typically find that the peak absorption amplitude is relatively constant between successive shots but decreases slightly over a few minutes, which we attribute to modification of the MTO pellet surface by the ablation laser that makes subsequent removal of material less efficient.  We find that the decay time of the absorption signal decreases much more significantly over this time, which we suggest is the result of changing helium density inside the cell --- we observe that the MTO pellet adsorbs significant amounts of helium when at low temperatures, and we speculate that a diminishing amount of this is liberated by successive ablation pulses.

At each laser frequency we average the data from 25 shots of the experiment.  By using a repetition rate of 12.5~Hz, phase-locked to the 50~Hz mains power frequency, we suppress electronic noise and ensure that the 25 shots are equally sampled over the vibrations induced by the 1~Hz mechanical motion of the cryocooler. In the time-dependent absorption profiles such as the one shown in figure~\ref{fgr:TOF}, we find that the peak absorption is more stable than the integrated absorption, and that the baseline varies greatly between shots primarily due to fringing effects caused by the windows which vibrate with the cooler. For these reasons, we determine the absorption at a given laser frequency by taking the standard deviation of the ordinate values in the averaged absorption signal.

Spectra are recorded by linearly stepping the laser current, $I_{\rm{QCL}}$, carrying out the 25 shot experimental sequence at each frequency.  At regular intervals we carry out wide scans covering up to 10~GHz to reconstruct a frequency scale. We divide up the etalon transmission signal into sections, typically 1-2 fringes wide, and fit Airy functions $A[f(I_{\rm{QCL}})]$ to each section, where $f(I_{\rm{QCL}})$ is a polynomial function. These fits establish the relationship between $I_{\rm{QCL}}$ and the relative frequency, allowing us to obtain a linear frequency scale with a relative uncertainty of a few megahertz. These wide scans also include a number of absorption lines from the reference gas cells that enable us to calibrate the etalon transmission function, giving us an absolute frequency accuracy of around 30~MHz. The accuracy and the relative uncertainty of the frequency scale are mainly limited by residual long-term thermal drifts of the etalon length which can be reduced with an improved temperature stabilisation. We use this periodically recalibrated frequency scale to carry out high-resolution scans over smaller regions that contain specific MTO absorption features of interest.

\begin{figure}
 \centering
 \includegraphics[width=\columnwidth]{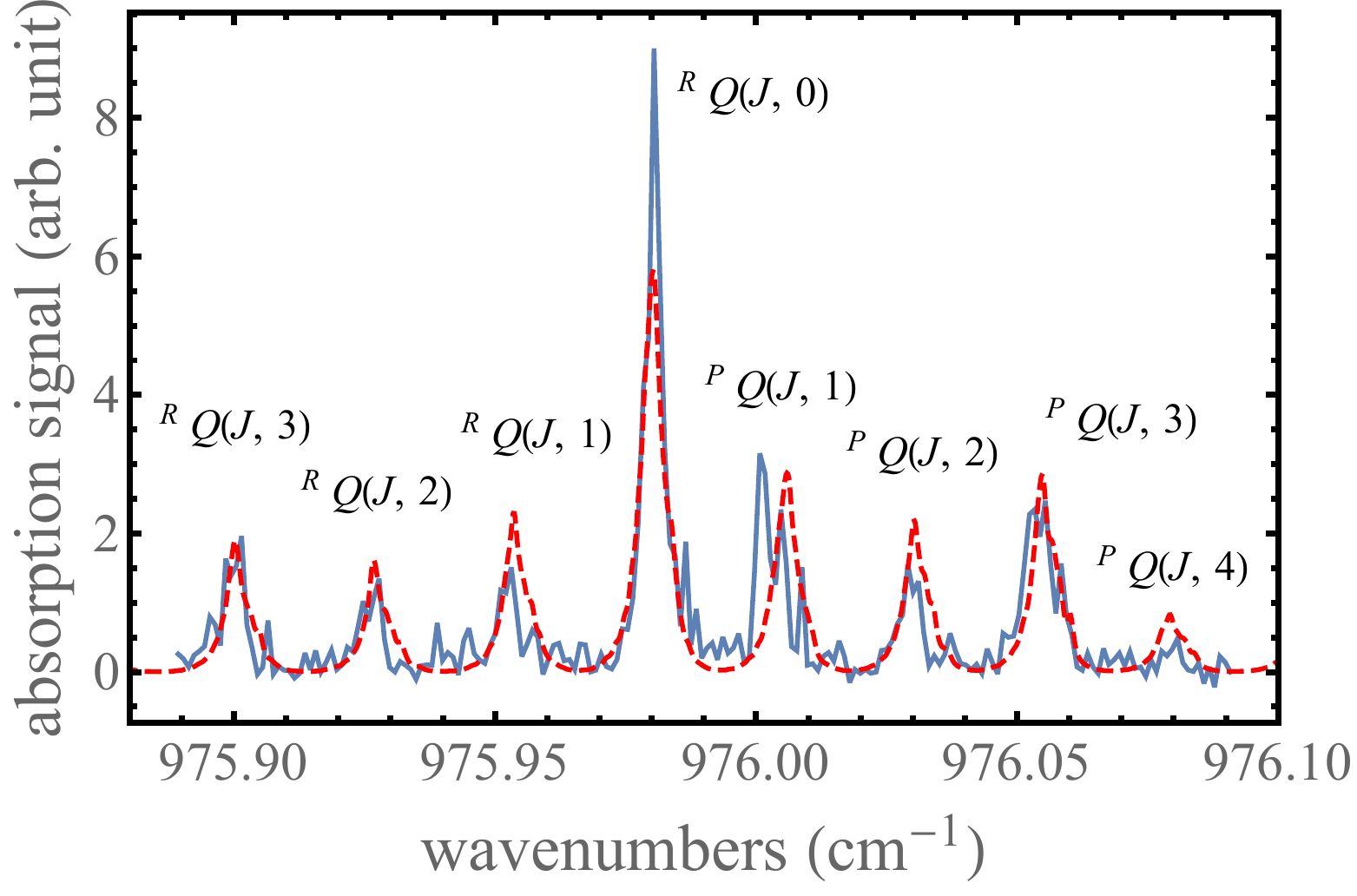}
 \caption{Linear absorption spectrum of the $Q$ branch of the antisymmetric Re=O stretching mode of the $^{187}$Re isotopologue of MTO. The height of the central peak corresponds to a few percent of absorption. The dashed (red) curve is a fit to the data (see text) from which a rotational temperature of $6\pm3$~K is inferred.}
 \label{fgr:spectra}
\end{figure}

Figure~\ref{fgr:spectra} shows an example spectrum. The observed structure corresponds to the rovibrational contour of the $Q$ branch of the $\nu_{\mathrm{as}}$ antisymmetric Re=O stretching mode\cite{Stoeffler2011} of the $^{187}$Re isotopologue of MTO centred around 976~cm$^{-1}$. Near its centre, this $\nu_{\mathrm{as}}$ perpendicular band consists of a series of $^{P}Q(J,K)$ and $^{R}Q(J,K)$ branches for which $K$ is fixed. In this notation, $K$ is the quantum number associated to the projection of the total orbital angular momentum onto the molecular symmetry axis, and the superscripts $P$ and $R$ refer to transitions with $\Delta K=-1$ and $\Delta K=+1$ respectively. Each branch is composed of a number of lines which are unresolved in the spectrum shown, each corresponding to a different value of $J$ (the total orbital angular momentum quantum number), with $J\geq K$. The figure also shows a fit of the data to a model calculated using a bar spectrum convolved with Lorentzian profiles. In this model, the line centres and intensities are found from an analysis of this vibrational mode of MTO that combines\cite{Asselin2016} microwave, millimetre-wave and infrared data, as well as the data presented in this paper. We use PGOPHER\cite{PGOPHER} to simulate spectra and fit the overall shape of the experimental data. We find that hyperfine structure has to be included in the model in order to accurately reproduce our spectra. We take the ground state hyperfine parameters obtained from microwave and millimeter-wave spectroscopy\cite{Asselin2016} and use these for both the ground and excited vibrational states. Once the line centres and relative intensities are fixed, the remaining free parameters are a vertical scaling, the rotational temperature, $T_{\mathrm{rot}}$, and the width of the individual Lorentzian profiles. We find $T_{\mathrm{rot}}=6\pm3$~K. Here, the uncertainty is dominated by a correlation between the vertical scaling factor and the rotational temperature, and it is determined from the deviation in $T_{\mathrm{rot}}$ values found after various modifications in the fitting procedure (choice of initial parameter guesses, convolution with a Lorentzian or a Gaussian).

\begin{figure*}
 \centering
 \includegraphics[width=\linewidth]{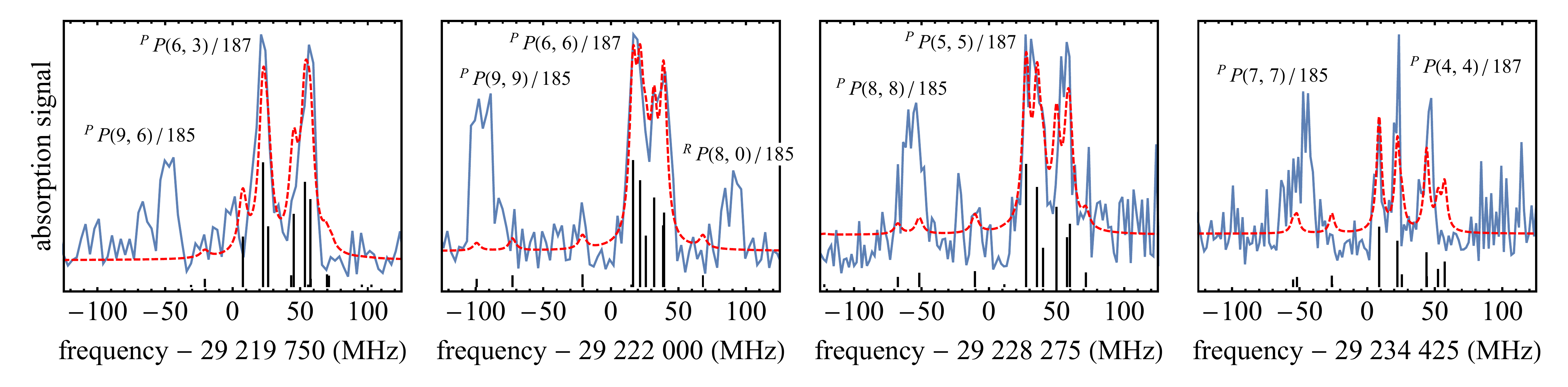}
 \caption{Linear absorption spectra in the $P$ branch of the antisymmetric Re=O stretching mode of both MTO isotopologues. The highest peaks correspond to a few percent absorption. The observed lines are labeled with transition assignment/rhenium atomic mass number. The dashed (red) curve is a fit to the data that takes into account the partially resolved hyperfine structure. The bar spectrum indicates the positions and relative intensities of the fitted hyperfine components. Note that the model spectrum does not include the 185 isotopologue.}
 \label{fgr:Pbranch}
\end{figure*}

Further to the red of the $Q$ branch, several $P$ lines of both the $^{185}$Re and $^{187}$Re isotopologues of MTO were also probed. Figure~\ref{fgr:Pbranch} shows four of these spectra exhibiting both $^{P}P(J,K)$ and $^{R}P(J,K)$ lines. Here, the splitting in $J$ is far bigger, and isolated rovibrational lines can thus be observed. Using these data, we measure the absolute frequencies of a number of transitions, as listed in table \ref{tbl:lineTable}. The quantum number assignments are determined with the help of data from Asselin {\it et al.}\cite{Asselin2016}. In the data shown in figure \ref{fgr:Pbranch}, we resolve neighbouring rovibrational lines of the $^{185}$Re and $^{187}$Re isotopologues of MTO. This is an improvement on two previous spectroscopic studies of supersonic jet-cooled MTO where the resolutions were 150~MHz (Fourier transform infrared spectroscopy\cite{Stoeffler2011}) and 120~MHz (laser absorption spectroscopy\cite{Asselin2016}). Thus, our data can improve the accuracy of some of the spectroscopic constants when used in a combined analysis with other datasets at various resolutions. That analysis will be presented elsewhere\cite{Asselin2016}. Here, we focus on an analysis of the hyperfine structure, which is partially resolved in our spectra.

\begin{table}
\small
  \caption{\ Centre frequencies of assigned rovibrational lines. The isotopologue is identified by the rhenium atomic mass number. Unresolved lines are assigned the same frequency. The 30~MHz uncertainty on transition frequencies is dominated by the long term drift of the reference etalon (see text)}
  \label{tbl:lineTable}
  \begin{tabular*}{0.48\textwidth}{@{\extracolsep{\fill}}lll}
    \hline
     Transition & Isotopologue & Frequency (THz) \\ \hline
     $^{P}P(9,6)$ & 185 & 29.219696(30) \\ \hline
     $^{P}P(6,3)$ & 187 & 29.219788(30) \\ \hline
     $^{P}P(9,9)$ & 185 & 29.221909(30) \\ \hline
     $^{P}P(6,6)$ & 187 & 29.222026(30) \\ \hline
     $^{R}P(8,0)$ & 185 & 29.222090(30) \\ \hline
     $^{R}P(5,0)$ & 187 & \multirow{2}{*}{29.224478(30)} \\ \cline{1-2}
     $^{P}P(8,3)$ & 185 & \\ \hline
     $^{P}P(8,8)$ & 185 & 29.228220(30) \\ \hline
     $^{P}P(5,5)$ & 187 & 29.228315(30) \\ \hline
     $^{R}P(4,0)$ & 187 & \multirow{2}{*}{29.231475(30)} \\ \cline{1-2}
     $^{P}P(7,3)$ & 185 & \\ \hline
     $^{P}P(7,7)$ & 185 & 29.234375(30) \\ \hline
     $^{P}P(4,4)$ & 187 & 29.234450(30) \\ \hline
  \end{tabular*}
\end{table}

The hyperfine structure is dominated by the interaction between the rhenium nuclear quadrupole moment and the gradient of the electric field at the nucleus. The spin\textendash rotation interaction between the magnetic field induced by the molecular rotation and the magnetic moment of the rhenium nucleus is much smaller, but large enough that we include it in our analysis\cite{Stoeffler2011,Asselin2016}. Smaller magnetic hyperfine structure resulting from the spins of hydrogen nuclei can be neglected. We also neglect matrix elements of the electric quadrupole Hamiltonian that are off-diagonal in $J$, since their inclusion shifts the transition hyperfine components by less than 170~kHz.

Since the rhenium nuclear spin is $I=5/2$ for both isotopes, each of the rovibrational levels corresponding to the transitions in figure~\ref{fgr:Pbranch} is split into six $F$ sub-levels, where $F$ is the quantum number of the operator $\mathbf{F}=\mathbf{J}+\mathbf{I}$.
The electric quadrupole interaction of vibrational state v is characterized by the quadrupole coupling constant $eQq^{\mathrm{v}}$. For a symmetric top like MTO, the energy shift is given to first order by\cite{Townes1955}
 \[ \Delta E^{\mathrm{v}}_{\mathrm{Q}}=eQq^{\mathrm{v}}\left( 3\frac{K^2}{J(J+1)}-1\right) f(J,I,F), \]
with $f(J,I,F)$ the Casimir function given by
 \[ f(J,I,F)=\frac{0.75 C(C+1)-I(I+1)J(J+1)}{2I(2I-1)(2J-1)(2J+3)}, \]
and
 \[ C=F(F+1)-I(I+1)-J(J+1). \]
The centrifugal distortion induced modification of the electric field gradient leads to a dependence of the quadrupole coupling constant on $J$ and $K$\cite{Spirko1979}
 \[ eQq^{\mathrm{v}}=eQq_0^{\mathrm{v}}+eQqJ^{\mathrm{v}}(J(J+1)-K^2). \]
The energy shift due to the nuclear spin-rotation magnetic interaction in the vibrational state v is characterized by the spin-rotation parameters $C_{\mathrm{aa}}^{\mathrm{v}}$ and $C_{\mathrm{bb}}^{\mathrm{v}}$ (because of the $C_{3\mathrm{v}}$ symmetry of MTO, $C_{\mathrm{bb}}^{\mathrm{v}}=C_{\mathrm{cc}}^{\mathrm{v}}$) and is given by\cite{Townes1955}
 \[ \Delta E^{\mathrm{v}}_{\mathrm{SR}}=\left(C_{\mathrm{bb}}^{\mathrm{v}}+(C_{\mathrm{aa}}^{\mathrm{v}}-C_{\mathrm{bb}}^{\mathrm{v}})\frac{K^2}{J(J+1)}\right)C. \]
The experimental spectra of the $^{P}P(6,3)$, $^{P}P(6,6)$, $^{P}P(5,5)$ and $^{P}P(4,4)$ lines of the $^{187}$Re MTO isotopologue displayed in figure~\ref{fgr:Pbranch} have been fitted by a bar spectrum convolved with Lorentzian profiles. Each of these transitions has six strong $\Delta F=\Delta J=-1$ hyperfine components and nine weaker $\Delta F=0,+1$ components. Their relative intensities, proportional to the square of the dipole moment matrix elements, are given by\cite{Thaddeus1964}
 \[ (2F^1+1)(2F^0+1)\left\{
     \begin{array}{ccc}
      J^1 & F^1 & I \\
      F^0 & J^0 & 1 \end{array}
     \right\}^2, \]
where $F^{\mathrm{v}}$ and $J^{\mathrm{v}}$ are the quantum numbers in the lower ($\mathrm{v}=0$) and upper ($\mathrm{v}=1$) states. There are two sets of constants $eQq_0^{\mathrm{v}}$, $eQqJ^{\mathrm{v}}$, $C_{\mathrm{aa}}^{\mathrm{v}}$ and $C_{\mathrm{bb}}^{\mathrm{v}}$ for the lower ($\mathrm{v}=0$) and the upper ($\mathrm{v}=1$) rovibrational levels. Those in $\mathrm{v}=0$ are fixed to the values extracted primarily from microwave and millimetre-wave spectroscopy\cite{Asselin2016}. The spin-rotation parameters in the excited state are too small for us to determine and so we fix them to their values in $\mathrm{v}=0$ ($C_{\mathrm{aa}}^1=C_{\mathrm{aa}}^0$ and $C_{\mathrm{bb}}^1=C_{\mathrm{bb}}^0$). The centrifugal distortion parameter in the excited state, $eQqJ^{\mathrm{1}}$, is also too small to determine, and we set it to zero. This leaves only the excited state quadrupole coupling constant $eQq^1=eQq_0^1$ which we float in the fitting procedure. Other adjustable parameters were, for each rovibrational line, the baseline offset, and the central frequency and width (identical for all the hyperfine components of a given line) of the Lorentzian. The four lines are fitted together as a single data set to obtain the best estimate of $eQq^1$. In these fits, we assign identical error bars to each point of a spectrum, deduced from the standard deviation of values measured in spectral regions where no signal is present.

Figure~\ref{fgr:Pbranch} shows the satisfactory agreement between the data and the fit. Note that the additional lines in fig~\ref{fgr:Pbranch} are signals from the $^{185}$Re isotopologue and are irrelevant to this discussion. We obtain $eQq^1=716(3)$~MHz, to be compared to $eQq^0=716.573(3)$~MHz in the ground vibrational state\cite{Asselin2016}, and conclude that there is little variation of the quadrupole coupling constant with vibrational quantum number. This situation is, for instance, different to that of ammonia (one of the few species whose hyperfine structure is known in both the ground and excited vibrational states) which exhibits 1 to 10\% $eQq$ variation depending on the vibration considered\cite{Lemarchand2011}. Note that the parameter uncertainty on $eQq^1$ reported by the fitting procedure is 1~MHz. We estimate a more conservative uncertainty from the range of values obtained after various modifications in the fitting procedure, including convolution with a Lorentzian or a Gaussian, fits with single or multiple spectra of the same line, and fits of only three among the four groups of lines.

From our fit, the full width at half maximum of individual hyperfine components is found to be $8\pm 2$~MHz. If we assume that this is entirely due to Doppler broadening, we deduce an upper limit to the translational temperature of $\sim35$~K. It seems unlikely that the translational temperature could be higher than the rotational temperature. Since the rotational temperature is consistent with the measured cell temperature, it seems most likely that the translation temperature is also around 6~K. In this case, the Doppler broadening contributes around $3$~MHz to the width. The remaining broadening may be due to laser frequency fluctuations and/or collisional broadening.

\section{Conclusion}

We have demonstrated the vibrational spectroscopy of an organometallic molecule cooled to low temperature in a helium buffer gas cell. The signal-to-noise ratio of our spectroscopy is similar to that of other techniques (jet-cooled Fourier transform infrared\cite{Stoeffler2011} or laser absorption fast-scan\cite{Asselin2016} spectroscopy), but has a resolution that is 20 times higher. This increase in resolution allows studies of excited vibrational state hyperfine structure, which is only known for a few polyatomic species and is certainly unprecedented for such a complex molecule. Higher resolution could be reached using saturated absorption spectroscopy in the buffer-gas cell once the quantum cascade laser is stabilised to an ultra-stable mid-infrared or near-infrared frequency reference as we recently demonstrated\cite{Sow2014,Argence2015}. Our work is a significant step towards producing an intense, slow-moving beam of cold chiral molecules for measurements of parity violation.

\section*{Acknowledgment}

The authors thank A. Amy-Klein, P. Asselin, C.J. Bord\'{e}, C. Chardonnet, C. Daussy, E.A. Hinds and T.R. Huet for fruitful discussions and L. Letouz\'{e} for the early characterization of the QCL. In France, this work was supported by ANR (under grants n\textsuperscript{o} ANR 2010 BLAN 724 3, n\textsuperscript{o} ANR-12-ASTR-0028-03 and n\textsuperscript{o} ANR-15-CE30-0005-01, and through Labex First-TF ANR 10 LABX 48 01), R\'{e}gion \^{I}le-de-France (DIM Nano-K), CNRS, Universit\'{e} Paris 13 and AS GRAM. In the UK, the work was supported by EPSRC under grant EP/I012044/1. The work was made possible through the International Exchanges Programme run jointly by the Royal Society and CNRS.

\bibliography{MTOBGHAL-arXivbib}

\end{document}